\documentclass[12pt,fleqn]{article}

\usepackage{t1enc}

\usepackage[latin1]{inputenc}

\usepackage{graphicx}

\usepackage[dvips]{rotating}

\usepackage{multicol}

\usepackage{flafter}

\usepackage{latexsym}

\usepackage{amssymb}

\begin{document}

{ \large \bfseries 3D Ising Nonuniversality: a Monte Carlo study}\\
\newline

{\large Melanie Schulte and Caroline Drope}\\
\newline
{\small Institute for Theoretical Physics, Cologne University,
50923 K\"{o}ln, Germany.

Email: ms@thp.uni-koeln.de, drope@thp.uni-koeln.de.}
\newline

\begin{abstract}
We investigate as a member of the Ising universality class the
Next-Nearest Neighbour Ising model without external field on
a simple cubic lattice by using the Monte Carlo Metropolis Algorithm.
The Binder cumulant and the susceptibility ratio, which should be universal
quantities at the critical point, were shown to vary for small negative
next-nearest neighbour interactions.
\end{abstract}

keywords: next-nearest neighbour Ising model, Binder cumulant,
fourth order cumulant.
\section{Introduction}
The great variety of critical behaviour is reduced by dividing all systems
into a small number of universality classes.
These classes are characterized by global properties like dimensionality and
the number of components of the order parameter and have the
same set of critical exponents and scaling functions, which are independent
of microscopic details \cite{landau}.
Here the 3D Ising universality class is considered by studying the
Binder cumulant \cite{binder} of the
Next-Nearest Neighbour Ising model (NNN model).
If all universal quantities of one class are known, the asymptotic
critical behaviour is believed to be also known completely under the condition,
that only two nonuniversal amplitudes $K_t$ and $K_h$ are
given. This particular feature is called two-scale factor
universality. The free-energy-density $f$, from which we can derive
thermodynamical quantities and which is measured in units of $k_B$, can be
expressed by \cite{privman}:
\begin{eqnarray*}
f(t,H;L)=f_s(t,H;L)+f_{ns}(t;L) \, ,
\end{eqnarray*}
where $t=1-T/T_c$ and $H$ is the ordering field. The singular part $f_s$ is the
part which yields thermodynamical singularities in the $L \rightarrow \infty$
limit. $f_{ns}$ denotes the non-singular part of $f$, which can be chosen
without field dependence.
For the singular part $f_s$ we can write for the Ising model
\begin{eqnarray*}
\fbox{$ \displaystyle f_s(t,H;L)=L^{-d} Y(K_t t L^{1/\nu},
K_h H L^{\delta/\nu})$}
\end{eqnarray*}
$Y(x,y)$ is called the scaling function and is universal using
system-dependent factors $K_h$ and $K_t$.\\
Chen and Dohm \cite{chen} recently predicted some deviations from this
universality, from a $\phi^4$ theory.\\
\newline
In the following section we study the universal value of the
Binder cumulant for the NNN model for different next-nearest neighbour
interaction $J_{NNN}$ by simulations. According to Privman, Hohenberg and
Aharony \cite{privman}, the Binder cumulant scales according to
\begin{eqnarray*}
U_4(t,L) \approx \overline{G}(K_t t L^{1/\nu}) \quad ,
\end{eqnarray*}
where $\overline{G}(x)=[(\partial^4 Y/ \partial y^4)/(\partial^2 Y/
\partial y^2)^2]_{y=0}$.
For $t=0$ (means $T=T_c$), $U_4(0,L)$ approaches with growing $L$ a universal
constant $U^*_4(0,\infty)$,
$U^*_4(0,L)$ which we are going to simulate, shows only small deviations
from the universal value $U^*_4(0,\infty)$ \cite{landau3}.
Consequently, it is possible to give a realistic approximation of the final
value $U^*_4(0,\infty)$ simulating finite lattices.

\section{Applied NNN Models}
\subsection{Standard Next-Nearest Neighbour Model}
To show the variation of $U^*_4$ within the 3D Ising universality class,
we choose the Next-Nearest Neighbour model (NNN Model) without external field
and a simple
cubic lattice. Each lattice site has 6 next neighbours in the distance $a$ and
12 next-nearest neighbours in the distance
$\sqrt{2} \cdot a$, if $a$ is defined as the lattice constant.
The exchange force $J_{NNN}$ of the next-nearest neighbours can now be
chosen to be different from the exchange force $J_{NN}$ of the 6
nearest neighbours as we also would also
intuitively expect, because their distance to the considered site is also
not identical.
\newline
A ferromagnetic phase transition is also existent for an antiferromagnetic
exchange force $J_{NNN}$. This kind of exchange, ferromagnetic for nearest
neighbours and antiferromagnetic for next-nearest neighbours, will be treated
for our universality test of $U^*_4$.

\subsection{Anisotropic Next-Nearest Neighbour Model}
Furthermore we studied the universality of $U_{4}^*$ in the zero field
NNN Ising Model
with an isotropic nearest-neighbour (NN) coupling $J>0$ and an anisotropic
next-nearest neighbour (ANNN) coupling $J_{NNN}<0$. We refer to the considered
model of Chen and Dohm in their publication ``Nonuniversal finite-size scaling
in anistropic systems`` [4]. The anisotropic NNN Ising Model is established by
considering only 6 of the 12 next-nearest neighbours being effective for NNN
interaction $J_{NNN}<0$ and the other 6 NNN having no NNN interaction. The
effective next-nearest neighbours have the positions
$\pm(1,0,0)$, $\pm(1,0,1)$ and $\pm(0,1,1)$ of  a simple cubic lattice.

\section{Simulation Results}
Simulating the NNN model $J_{NNN}$ can be varied, while it is chosen
$J_{NN}=1$.
For values of $J_{NNN}$ higher than a critical value
$J_{NNN,crit} \approx -0.5$ we observe ferromagnetic phase transitions, while
below $J_{NNN,crit}$ no ferromagnetic phase transition occurs. At $J_{NNN}=0$
we obtain the standard 3-dimensional (3D) Ising ferromagnet. Here the region
$J_{NNN,crit}<J_{NNN}<0$ and thus a small antiferromagnetic interaction
$J_{NNN}$ is considered.\\
Simulating now the "fixed point value" at the critical point of the
Binder cumulant $U^*_4=U_4(T_c)$, we evaluate $U^*_4(J_{NNN})$
to check its universality within the 3D Ising universality class.
For an Ising Model in zero field, the Binder cumulant simplifies to
\begin{equation}
U_{4}=1-\frac{\left< M^4 \right>}{3\left< M^2 \right>^2}\, \ \, ,
\end{equation}
where $M$ symbols the magnetization per spin.
For the determination of each $U^*_4$ was plotted $U_4$ as a function of
temperature $T$ for three different lattice lengths $L$ ($=10$, $20$, $40$).
To get more precise results,
several curves of $U^*_4(T)$ for different lattice sizes with different seeds
were
averaged. In figure 1 is shown this method
for two different $J_{NNN}$. The errors were calculated by averaging the values
of the three different crossing points of $L=(10,20)$, $(10,40)$, $(20,40)$.
\begin{figure}[h]
\includegraphics[width=6.0cm,angle=-90]{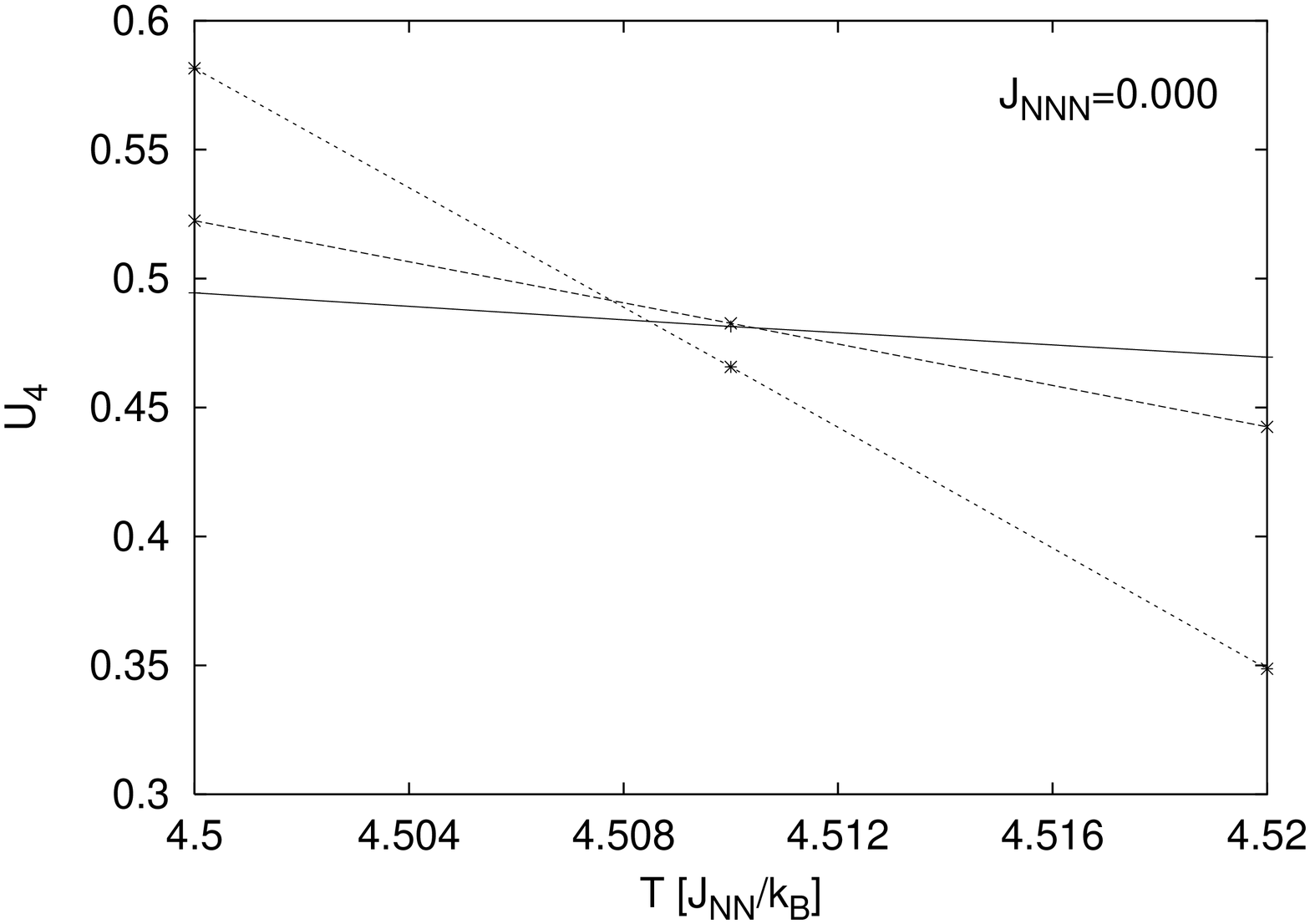}
\includegraphics[width=6.0cm,angle=-90]{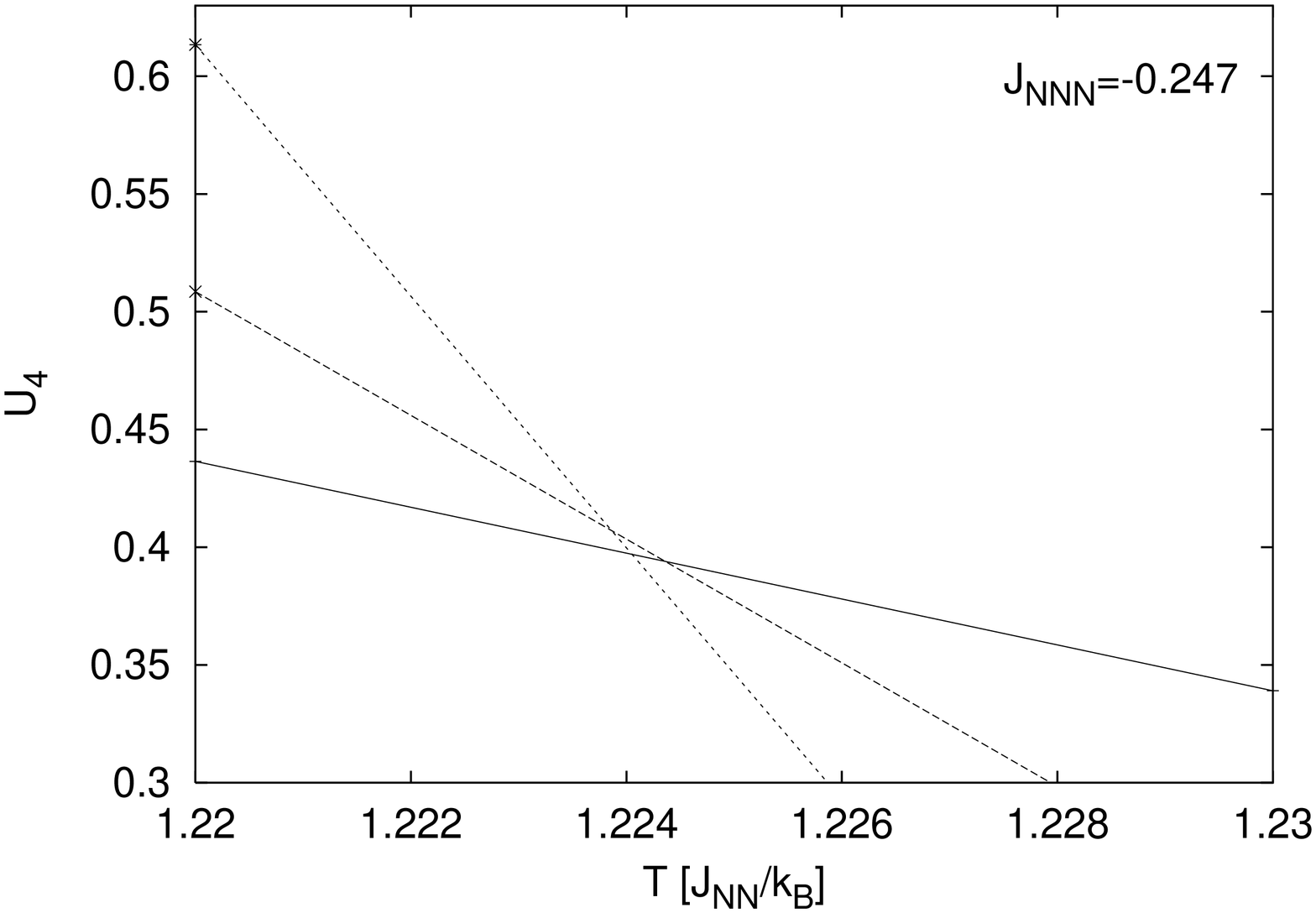}
\label{sdf1}
\caption{Localization of the fixed point ($U^*_4$,$T_c$) for different
$J_{NNN}$ by quantifying the crossing point of $U_4(T)$ for lattice sizes
$L=10$ (dotted), $20$ (dashed) and $40$ (solid).}
\end{figure}

\subsection{Results for the Standard NNN Ising Model}
\label{nnnresults}
\begin{figure}[h]
\includegraphics[width=6.0cm,angle=-90]{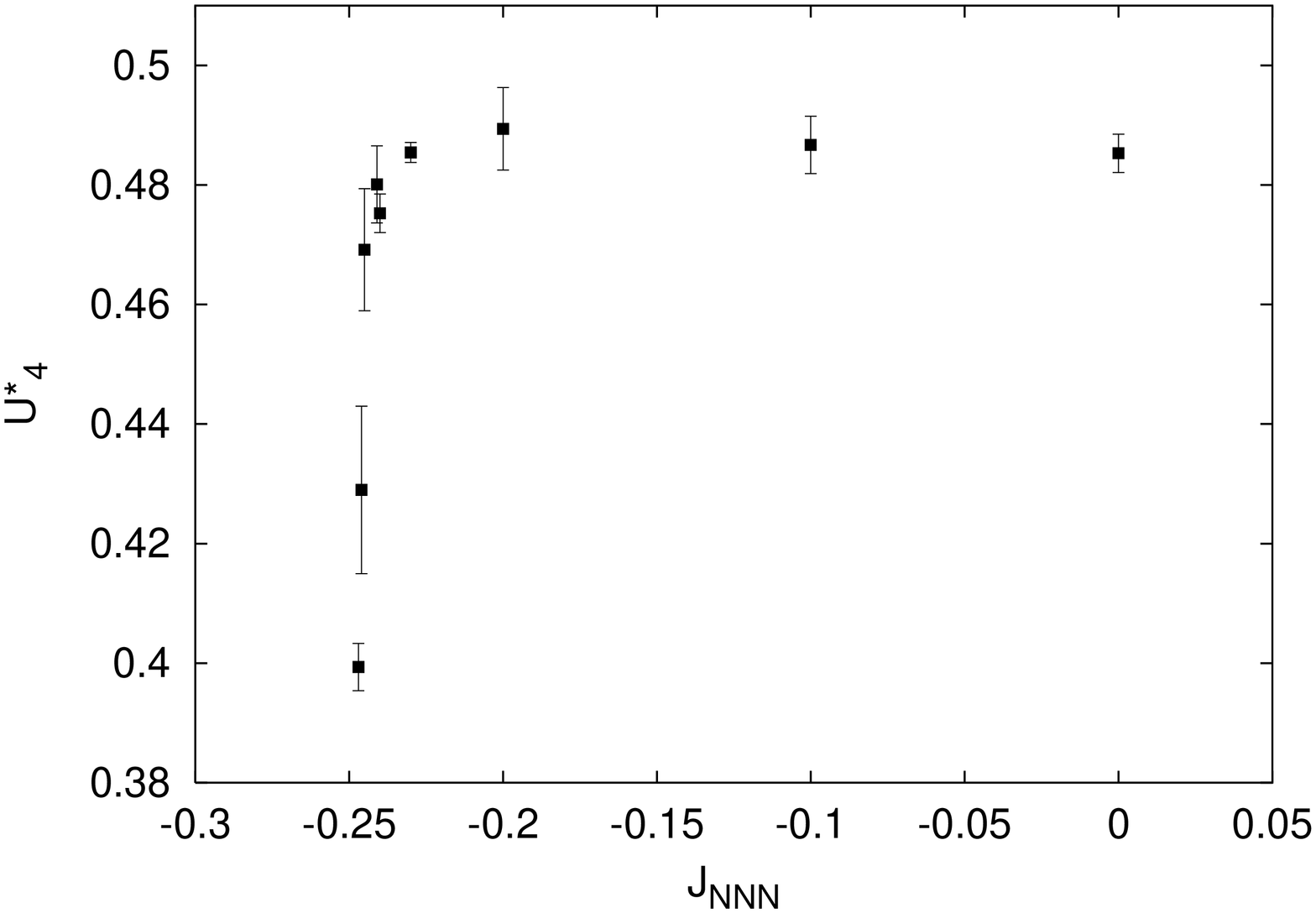}
\includegraphics[width=6.0cm,angle=-90]{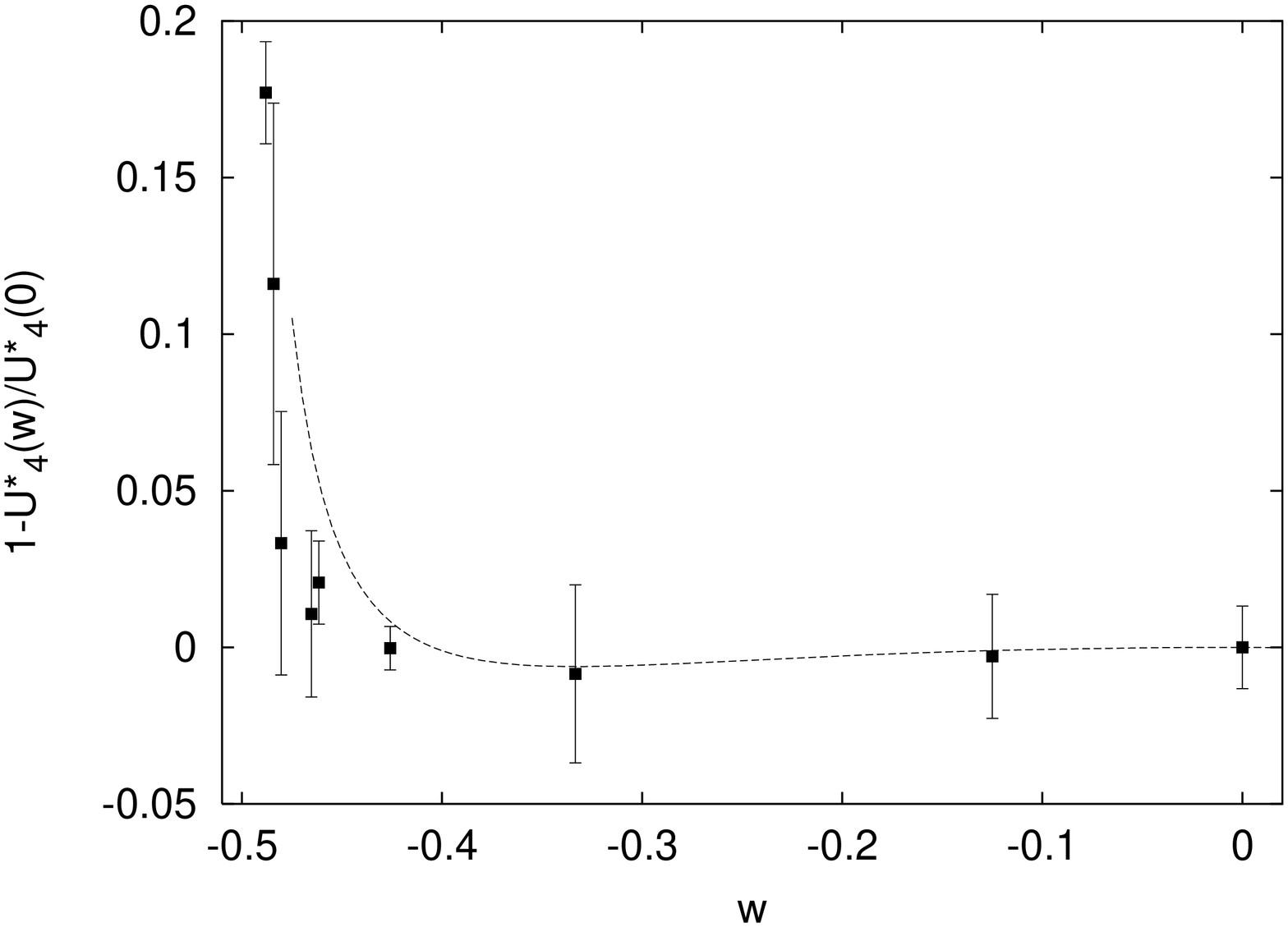}
\caption{Standard 3D NNN model. Left side: Simulation results of the fixed
point $U^*_4$ as a function of $J_{NNN}$; Right side: Results of the figure
on the left
({\tiny $\blacksquare$}) in comparison with theoretical results (for a different
neighbourhood) of Chen and Dohm \cite{chen} (dashed line).}
\label{ende}
\end{figure}

In figure \ref{ende} and table 1 we can see all simulated
results as a function $U^*_4(J_{NNN})$ and these results in comparison with
theoretical results of Chen and Dohm \cite{chen}
plotted as a function $1-U^*_4(w)/U(0)$ with
$w=J_{NNN}/(J_{NN}+2J_{NNN})$.
Chen and Dohm considered an anisotropic NNN Ising model, which we will
discuss later.
\begin{table}
\centering
\begin{tabular}{|l|l||l|l|} \hline
{\bfseries $J_{NNN}$}&{\bfseries $w$} &{ \bfseries $U^*_4$}&{ \bfseries $1-U^*_4(w)/U^*_4(0)$} \\ \hline\hline
$\,\,\,\,\,0.000$&$\,\,\,\,\,0.000$&$0.4853\pm3.20\cdot10^{-3}$&$\,\,\,\,\,0.0000\quad\quad\,\,\,\,\pm1.32\cdot10^{-2}$\\ \hline
$-0.100$&$-0.125$&$0.4867\pm4.80\cdot10^{-3}$&$-2.8609\cdot10^{-3}\pm1.98\cdot10^{-2}$\\ \hline
$-0.200$&$-0.333$&$0.4894\pm6.90\cdot10^{-3}$&$-8.4366\cdot10^{-3}\pm2.84\cdot10^{-2}$\\ \hline
$-0.230$&$-0.426$&$0.4854\pm1.68\cdot10^{-3}$&$-2.6201\cdot10^{-4}\pm6.92\cdot10^{-3}$\\ \hline
$-0.240$&$-0.462$&$0.4753\pm3.22\cdot10^{-3}$&$\,\,\,\,\,2.0685\cdot10^{-2}\pm1.33\cdot10^{-2}$\\ \hline
$-0.241$&$-0.465$&$0.4801\pm6.44\cdot10^{-3}$&$\,\,\,\,\,1.0700\cdot10^{-2}\pm2.65\cdot10^{-2}$\\ \hline
$-0.245$&$-0.480$&$0.4692\pm1.02\cdot10^{-2}$&$\,\,\,\,\,3.3247\cdot10^{-2}\pm4.20\cdot10^{-2}$ \\ \hline
$-0.246$&$-0.484$&$0.4290\pm1.40\cdot10^{-2}$&$\,\,\,\,\,0.1161\quad\quad\,\,\,\,\pm5.77\cdot10^{-2}$ \\ \hline
$-0.247$&$-0.488$&$0.3994\pm3.96\cdot10^{-3}$&$\,\,\,\,\,0.1771\quad\quad\,\,\,\,\pm1.63\cdot10^{-2}$ \\ \hline
\end{tabular}\\
\caption{Simulation results of the universal quantity $U^*_4$ and its
statistical errors for different next-nearest neighbour interaction $J_{NNN}$.}
\end{table}
It clearly can be seen in figure and, that $U^*_4$
is {\it not} a fixed value for different $J_{NNN}$, particularly
in the interval of $J_{NNN}=$ [-0.25:-0.22]. This implies in this region the
{\it absence of universality} for the 3D Standard NNN Ising model.\\
Great deviations of the resulting $U^*_4$ for $L \rightarrow \infty$
can be excluded for $J_{NNN}=0$ \cite{landau3}.
Observing lattice length up to $L=100$ for $J_{NNN}\ne 0$ we observed no
significant change of our simulation results.\\
We can conclude, that in the region of small negative values for
$J_{NNN}$ the 3D Next-Nearest Neighbour Ising model shows a different universal
behaviour than the 3D Nearest Neighbour Ising model.

\subsection{Results for the ANNN Ising Model}
\begin{figure}
\includegraphics[width=6.0cm,angle=-90]{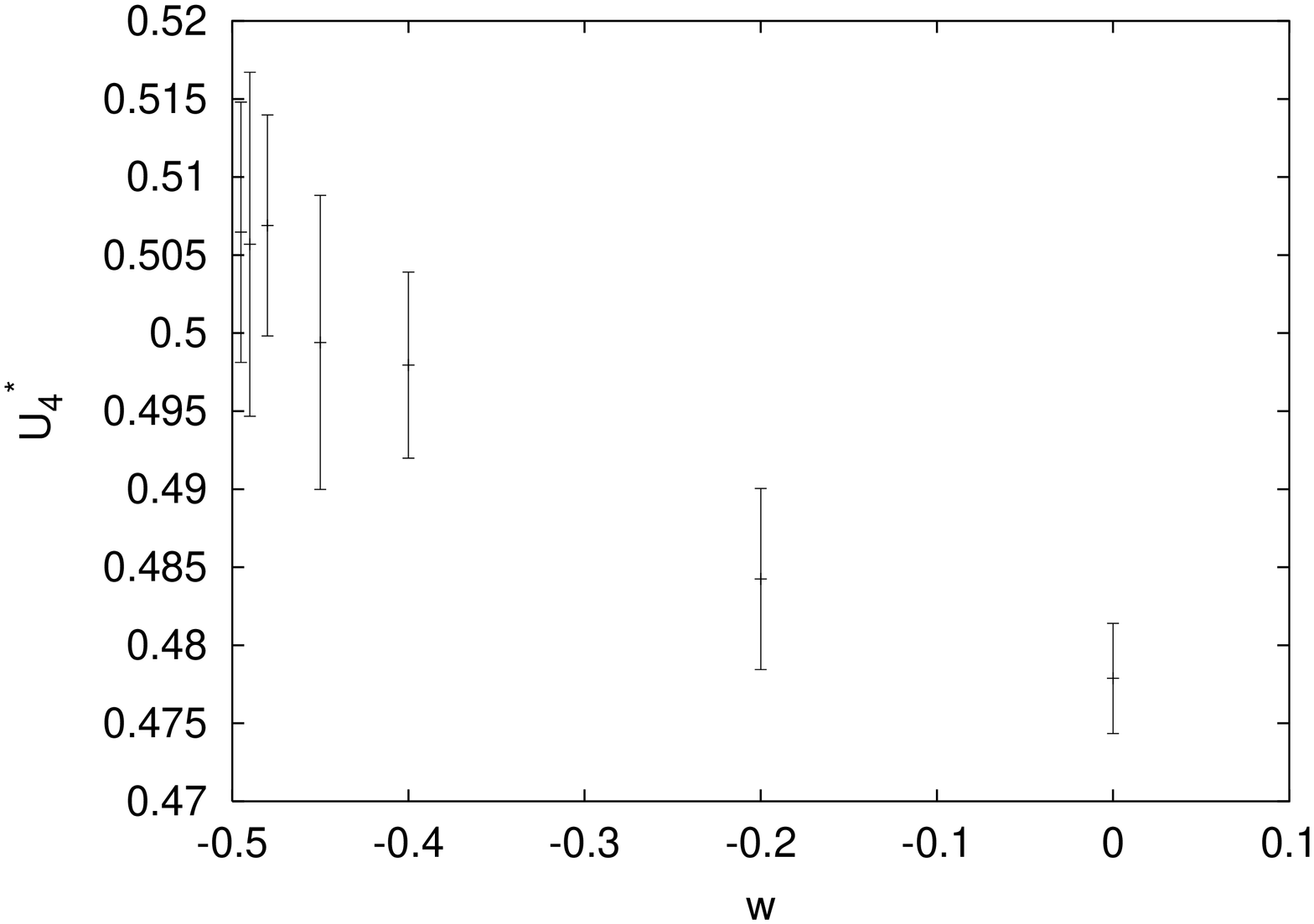}
\includegraphics[width=6.0cm,angle=-90]{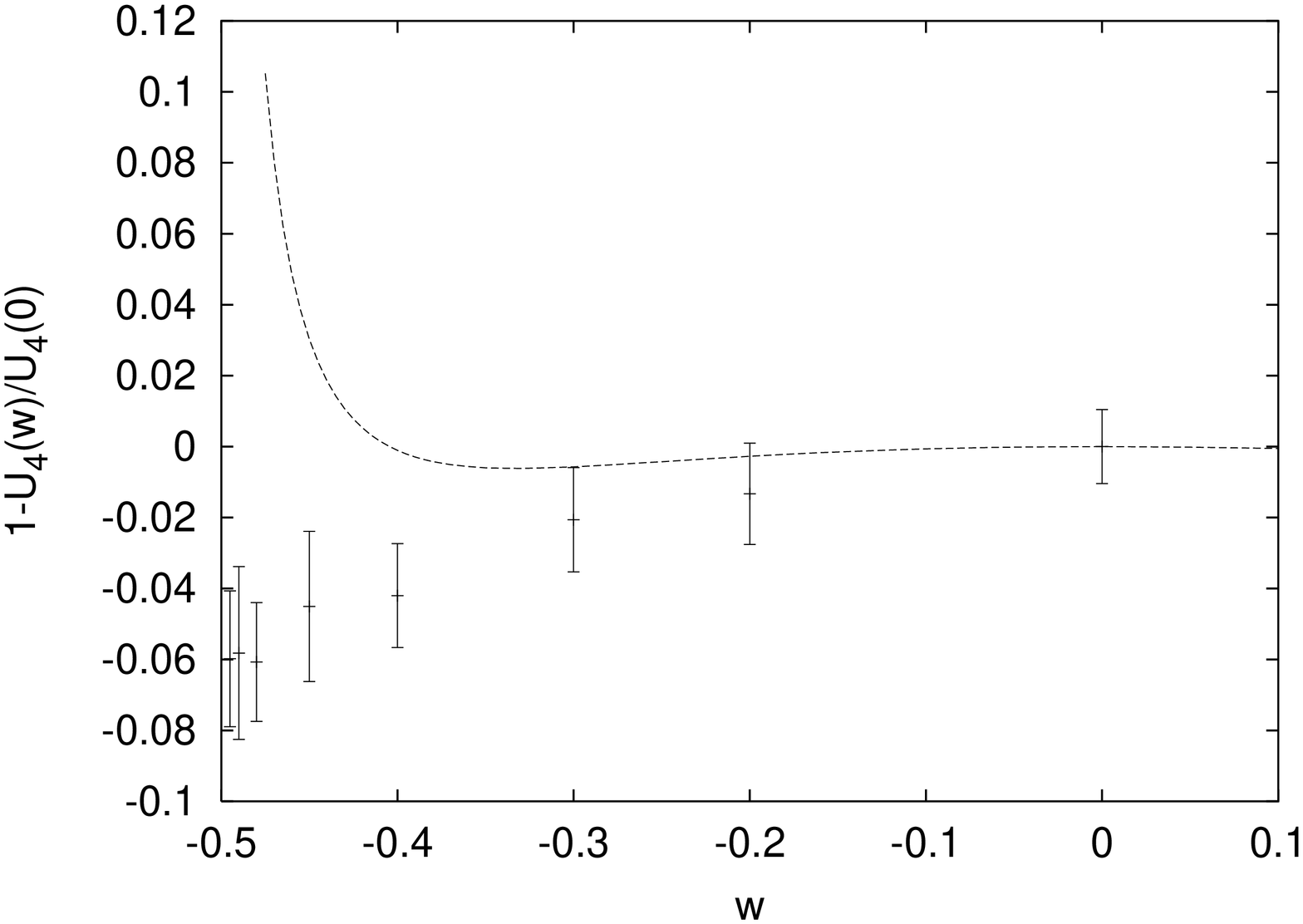}
\caption{Left side: Simulation results ({\small $+$}) for the ANNN model of $U^*_4(w)$.
Right side: Simulation results of $1-{U_{4}^*(w) / U_{4}^*(0)}$ in comparison
with the theoretical results (for this neighbourhood) of Chen and Dohm
\cite{chen} (solid line).}
\label{caro}
\end{figure}

Figure \ref{caro} and table 2 show the results for the ANNN Ising Model analogue to section
\ref{nnnresults}. The simulation results show a significant non-universality
of the Binder cumulant. As compared to the analytic results of Chen and Dohm
\cite{chen} the analytical curve have a different tendency and is not within
the errorbars
of the simulation results. For $w \to -0.5$,  $U_{4}^*$ increases whereas Chen
and Dohm obtained a decrease. Thus, the theoretical results cannot be
confirmed fully.
\begin{table}
\centering
\begin{tabular}{|c|c|c|c|}
\hline
$J_{NNN}$  &$w$     &$U_{4}^*$   & $1-{U_{4}^*(w) / U_{4}^*(0)}$\\

\hline\hline
$0$  &$0$     &$0.478 \pm 3.53 \cdot 10^{-3}$    &$0 \pm 1.04 \cdot 10^{-2}$\\
\hline
$-0.143$   &$-0.2$  &$0.484 \pm 5.80 \cdot 10^{-3}$    &$-1.33 \cdot 10^{-2} \pm 1.43\cdot 10^{-2}$ \\

\hline
$-0.188$    &$-0.3$  &$0.488 \pm 6.04 \cdot 10^{-3}$    &$-2.06 \cdot 10^{-2} \pm 1.47 \cdot 10^{-2} $\\
\hline
$-0.222$  & $-0.4$ &$0.498 \pm 5.96 \cdot 10^{-3}$   &$-4.20 \cdot 10^{-2} \pm 1.47\cdot 10^{-2}$\\
\hline
$-0.237$  &$-0.45$  &$0.499 \pm 9.42 \cdot 10^{-3}$   &$-4.50 \cdot 10^{-2} \pm 2.12\cdot 10^{-2}$\\
\hline
$-0.245$   &$-0.48$  &$0.507 \pm 7.08 \cdot 10^{-3}$   &$-6.07 \cdot 10^{-2} \pm 1.68\cdot 10^{-2}$\\
\hline
$-0.247$   &$-0.49$  &$0.507 \pm 1.10 \cdot 10^{-2}$   &$-5.82 \cdot 10^{-2} \pm 2.43\cdot 10^{-2}$\\
\hline
$-0.249$   &$-0.495$  &$0.507 \pm 8.34 \cdot 10^{-3}$   &$-5.98 \cdot 10^{-2} \pm 1.91\cdot 10^{-2}$\\
\hline
\end{tabular}
\caption{Simulation results for the ANNN Model for $U_{4}^*(w)$.}
\end{table}

\clearpage

\subsection{Universality Test of the Susceptibility for the ANNN Model}

\begin{figure}[ht]
\centering
\includegraphics[height=8cm, angle=270]{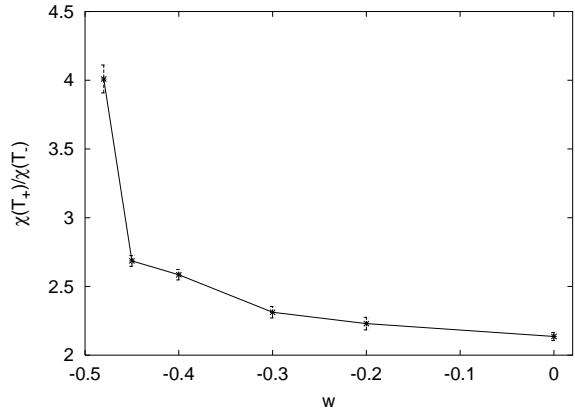}
\caption{Simulation results of the ratio of susceptibility ${\chi(T_{+})}/{\chi(T_{-})}$ as a
function of $J_{NNN}$.}
\label{sus+-}
\end{figure}

Simulating the susceptibility
$\chi=\left({N}/{k_BT}\right) \left( \left< M^2 \right> - \left<|M|\right>^2 \right)$
for the ANNN Ising Model
we use the same range of NNN interaction $J_{NNN}$.
We consider two temperatures
near the critical temperature $T_c$, $T_{-}=T_c - 3\% T_c$ and
$T_{+}=T_c + 3\% T_c$. The results show explicitly the absence of universality.
Figure \ref{sus+-} shows the ratio of the susceptibility at $T_+$ and $T_-$
as a function of $w$.
A significant deviation of nearly 100\% exists here between $w=0$ and
$w=-0.45$. Varying below $T_c$ the lattice size $L$ between $10$ and $317$
and the number of Monte Carlo steps $MCS$ between $50$ and $5 \cdot 10^5$
we can
exclude finite-size effects with $L=40$ and $MCS=5 \cdot 10^5$. Table \ref{sus}
 shows
all results of $\chi(T_+)$, $\chi(T_-)$ and ${\chi(T_+)}/{\chi(T_-)}$ for
different $J_{NNN}$. (We looked at the fluctuations in the absolute value
of the magnetization $<M^2> - <|M|>^2$, not of the magnetization. Thus our
values above $T_c$ are lower by a factor $1 - 2/\pi = 0.3634$ than the usual
$<M^2>$, assuming Gaussian fluctuations \cite{binder}.)

\begin{table}
\centering
\begin{tabular}{|c|c||c|c|c|} \hline
{\bfseries $J_{NNN}$}&{\bfseries $w$} &{ \bfseries $\chi(T_+)$}&{ \bfseries $\chi(T_-)$}&{ \bfseries ${\chi(T_+)}/{\chi(T_-)}$} \\ \hline\hline
$0$&$0$&$32.10\pm0.39$&$15.03\pm9.3\cdot10^{-2}$&$2.14\pm2.9\cdot10^{-2}$\\ \hline
$-0.143$&$-0.2$&$34.26\pm0.70$&$15.37\pm4.1\cdot10^{-2}$&$2.23\pm4.6\cdot10^{-2}$\\ \hline
$-0.188$&$-0.3$&$35.46\pm0.59$&$15.34\pm11.1\cdot10^{-2}$&$2.31\pm4.2\cdot10^{-2}$\\ \hline
$-0.222$&$-0.4$&$38.54\pm0.47$&$14.91\pm12.2\cdot10^{-2}$&$2.59\pm3.8\cdot10^{-2}$\\ \hline
$-0.237$&$-0.45$&$39.53\pm0.55$&$14.72\pm6.6\cdot10^{-2}$&$2.69\pm3.9\cdot10^{-2}$\\ \hline
$-0.245$&$-0.48$&$57.27\pm1.36$&$14.28\pm12.5\cdot10^{-2}$&$4.01\pm10.2\cdot10^{-2}$\\ \hline
\end{tabular}\\
\caption{Simulation results of the susceptibility at $T_{-}=T_c - 3\% T_c$ and
$T_{+}=T_c + 3\% T_c$ and of the ratio ${\chi(T_{+})}/{\chi(T_{-})}$ as a
function of $J_{NNN}$.}
\label{sus}
\end{table}

\clearpage

\section{Conclusions}
This work examined the universality of the Binder cumulant
for the 3D NNN Ising model in the region of small negative values for the
next-nearest neighbour interaction.
Monte Carlo simulations of the universal quantity of the
Binder cumulant and the universal susceptibility ratio with
different next-nearest neighbour
interactions $J_{NNN}$ showed a variation of both quantities for all applied
models.
The variation of these universal quantities leads to the conclusion, that in
the
considered region the NNN Ising model does not belong fully to the 3D Ising
universality class.\\
We assume that for regions, where chosen values of $J_{NN}$ and
$J_{NNN}$ still cause ferromagnetic phase transitions, but are close to a
region where we will find no more ferromagnetic behaviour,
a new definition for universality classes for this model has to be found.

\section{Acknowledgements}
We thank Dietrich Stauffer for contributing his knowledge and experiences
in Monte Carlo simulations.
Special thanks also to Volker Dohm from the technical university of
Aachen and Xiaosong Chen from the Institute of Theoretical Physics, Chinese
Academy of Sciences for their discussions and help.


\end{document}